# Anisotropic, Intermediate Coupling Superconductivity in $Cu_{0.03}TaS_2$

Xiangde Zhu[1], Yuping Sun[1,2,*], Shuhua Zhang[1], Jianglong Wang[3], Liangjian Zou[1,*], Lance E DeLong,[4] Xuebin Zhu[1], Xuan Luo[1], Bosen Wang[1], Gang Li[1], Zhaorong Yang[1], Wenhai Song[1]

[1] Key Laboratory of Materials Physics, Institute of Solid State Physics, Chinese Academy of Sciences, Hefei 230031, P. R. China

[2] High Magnetic Field Laboratory, Chinese Academy of Sciences, 230031, P. R. China

[3] College of Physics Science & Technology, Hebei University, Baoding, 071002, P. R. China

[4] Department of Physics and Astronomy, University of Kentucky, Lexington, KY 40506-0055 USA

Email: experiment and data analysis: ypsun@issp.ac.cn (Y. P. Sun);

theory: zou@theory.issp.ac.cn (L. J. Zou).

**Abstract**

The anisotropic superconducting state properties in $Cu_{0.03}TaS_2$ have been investigated by magnetization, magnetoresistance, and specific heat measurements. It clearly shows that $Cu_{0.03}TaS_2$ undergoes a superconducting transition at $T_C$ = 4.03 K. The obtained superconducting parameters demonstrate that $Cu_{0.03}TaS_2$ is an anisotropic type-II superconductor. Combining specific heat jump $\Delta C/\gamma_n T_C$ = 1.6(4), gap ratio $2\Delta/k_B T_C$ = 4.0(9) and the estimated electron-phonon coupling constant $\lambda \sim 0.68$, the superconductivity in $Cu_{0.03}TaS_2$ is explained within the intermediate coupling BCS scenario. First-principles electronic structure calculations suggest that copper intercalation of 2$H$-$TaS_2$ causes a considerable increase of the Fermi surface volume and the carrier density, which suppresses the CDW fluctuation and favors the raise of $T_C$.

# 1. Introduction

The compounds $2H$-$TaS_2$, $2H$-$TaSe_2$, and $1T$-$TiS_2$ are layered transition-metal dichalcogenides (TMDC) [1-2], formed by stacking covalently bonded X-T-X layers that are weakly coupled by van der Waals bonding. Additional atoms and organic molecules can be inserted into the gap between the layers, forming intercalated compounds. Since the discoveries of superconductivities in potassium intercalated graphite $KC_8$ in 1965 [3] and $(Py)_{1/2}TaS_2$ in 1970 [4], superconductivity induced by intercalation has been widely investigated in highly oriented pyrolitic graphite (HOPG) [5] and layered TMDC [6]. The recent discoveries of superconductivity in $CaC_6$ with $T_C$ = 11.5 K [7] and the suppression of charge density wave (CDW) order in $Na_xTaS_2$ [8], $Cu_xTiSe_2$ [9] have renewed interest in intercalated layered compounds. To date, the influence of transition-metal intercalation on the electronic structure is still not clear. For example, a rigid band model with charge transfer was proposed by assuming that intercalation only alters the density of state (DOS) at the Fermi level ($E_F$) without any other change in the electronic structure [10]. However, previous experiments in intercalated graphite compounds did not support the rigid band model [11]. It is therefore of interest to determine if the rigid band model is applicable to TMDC, which is a question not yet answered in the literature.

$2H$-$TaS_2$ is a typical layered TMDC that exhibits coexisting CDW (TCDW ~ 78K) and superconducting ($T_C$ ~ 0.8 K) phases [12]. Enhanced superconductivities have been discovered in many organic molecules intercalated $2H$-$TaS_2$ [4, 13, 14], alkali metal intercalated $A_xTaS_2$ [15], and 3d-transitonal metal Fe dilute intercalated $2H$-$Fe_{0.05}TaS_2$ [16]. Recently, the basic superconducting properties of the polycrystalline $Cu_xTaS_2$ have been reported by Cava group [17]. However, the anisotropic superconducting state parameters have not been characterized yet, and the origin of the $T_C$ enhancement by copper intercalation is still unknown. In this paper, we present the anisotropic superconducting properties of $Cu_{0.03}TaS_2$ single crystal and comprehensively analyze the superconductivity in $Cu_{0.03}TaS_2$ within the intermediate-coupling BCS scenario. We perform first-principles electronic structure calculations based on the tight-binding linear Muffin-Tin orbital (TB-LMTO) approach, suggesting that the substantial increase of $T_C$ in $Cu_{0.03}TaS_2$ originates from the enlargement of the Fermi surface volume and the increase of carrier density with copper intercalation, and that the significant alteration of the band structures near $E_F$ invalidates the rigid band approximation.

**2. Experimental details**

Dilute copper intercalates of composition $Cu_xTaS_2$ ($x$ = 0.03) were grown as single crystals via chemical vapor transport with iodine as a transport agent [18]. The magnetization ($M$) measurements were performed with a SQUID magnetometer (Quantum Design MPMS). The specific heat ($C$) and the magnetoresistance measurements were carried out with a physical property measurement system (Quantum Design PPMS).

The microscopic electronic states and possible origin of the profound increase in the superconducting temperature in $Cu_xTaS_2$ were investigated by the first-principles electronic structures calculations, utilizing the tight-binding linear Muffin-Tin orbitals within the atomic sphere approximation (TB-LMTO-ASA) scheme for the supercells of 2×2×1 and 2×2×2, corresponding to the Cu concentrations of x = 1/8 and 1/16.

**3. Experimental results and analysis**

*3.1 Anisotropic superconducting parameters*

Figure 1(a) shows the temperature dependence of the lower critical field $H_{C1}$ for **H** ∥ **ab** and **H** ∥ **c** (after demagnetization correction) determined from the measured *M-H* curves. The temperature dependence of $H_{C1}^i(T)$ (where $i$ donates the field applied along the $i$ direction) can be well fitted to $H_{C1}^i(T) = H_{C1}^i(0)[1-(T/T_C)^2]$ [19]. The inset of figure 1(a) shows the typical field dependence of *M-H* curves measured at $T$ = 2.2 K. Due to the plate-shape of the single-crystal sample, the demagnetization effect for **H** ∥ **ab** is negligible, while the demagnetization for **H** ∥ **c** is large. Demagnetization corrections of the *M-H* curves for **H** ∥ **c** were performed according to Ref. [20]. For comparison, the corrected *M-H* curve for **H** ∥ **c** is also shown in the inset of figure 1(a).

The upper critical field $H_{C2}$ can be obtained from the in-plane magnetoresistance measurements ($\rho_{ab}$-*H*) for **H** ∥ **ab** and **H** ∥ **c**. Figure 1(b) shows the typical temperature dependences of the upper critical field ($H_{C2}$) for **H** ∥ **ab** and **H** ∥ **c**, and the inset of figure 1(b) depicts $\rho_{ab}$-$H$ curves for **H** ∥ **ab** and **H** ∥ **c** at $T$ = 3.8 K. The $H_{C2}$(T) data exhibit an almost linear temperature dependence near $T_C$, which is consistent with the Werthamer-Helfand-Hohenberg (WHH) model for type II superconductors [21]. Extrapolations of $H_{C2}$(0) were performed with the WHH equation

$$H_{C2}(0) = 0.693[-(dH_{C2}/dT)]_{T_C} T_C .$$

The determined $H_{C1}$(0) and $H_{C2}$(0) are: $H_{C1}^{ab}(0) \approx 40$ Oe, $H_{C1}^{c}(0) \approx 135$ Oe, $H_{C2}^{ab}(0) \approx 9.16$ T, and $H_{C2}^{c}(0) \approx 1.8$ T.

According to the $H_{C2}$-$T$ relations in figure 1(b), the Ginzburg-Landau (GL) anisotropy parameter, $\gamma_{anis} = H_{C2}^{ab}/H_{C2}^{c} = 5.1$, is roughly temperature independent. with the GL formulas for anisotropic upper critical fields: $H_{C2}^{ab}(0) = \Phi_0/(2\pi\xi_{ab}\xi_c)$ and $H_{C2}^{c}(0) = \Phi_0/(2\pi\xi_{ab}^2)$, where $\Phi_0$ is the flux quantum, the GL coherence lengths are estimated to be $\xi_{ab}$ = 13.5 nm and $\xi_c$ = 2.65 nm, respectively. The GL parameters $\kappa_i(0)$ are also obtained with the equation $H_{C2}^{i}(0)/H_{C1}^{i}(0) = 2\kappa_i^2(0)/\ln\kappa_i(0)$. With $H_C(0) = H_{C1}^{ab}(0)/\sqrt{2}\kappa_{ab}(0)$, the thermodynamic critical field $H_C$(0) is determined to be ≈ 0.1 T. The GL penetration length is evaluated through $\kappa_c(0) = \lambda_{ab}(0)/\xi_{ab}(0)$ and $\kappa_{ab}(0) = \lambda_{ab}(0)/\xi_c(0) = [\lambda_{ab}(0)\lambda_c(0)/\xi_{ab}(0)\xi_c(0)]^{1/2}$. We note that the obtained superconducting parameters can be approximately fitted to the anisotropic GL relation

$$\gamma_{anis} = \frac{H_{C2}^{ab}}{H_{C2}^{c}} = \frac{\xi_{ab}}{\xi_c} = \frac{\lambda_c}{\lambda_{ab}} = \frac{\kappa_{ab}}{\kappa_c} \sim \frac{H_{C1}^{c}}{H_{C1}^{ab}} .$$

All these parameters are summarized in table 1, and they indicate that $Cu_{0.03}TaS_2$ is a typical type II superconductor with a large anisotropy.

*3.2 Specific results and discussions*

Data for the specific heat divided by temperature, $C/T$, for $Cu_{0.03}TaS_2$ are shown in figure 2 as a function of $T^2$ with magnetic fields $H$ = 0 and 2.5 kOe applied perpendicular to the **ab**-plane. The sharp jump in the specific heat data at $T_C$ = 4.03 K indicates the bulk nature of superconductivity and high quality of our $Cu_{0.03}TaS_2$ samples, which is corroborated by a sharp drop of the magnetic susceptibility (shown in the inset of figure 2) at $T$ = 4.2 K with a transition width (10%-90%) of 0.2 K, and a zero-field-cooling (ZFC) curve that indicates a perfect shielding effect.

The low-temperature specific heat $C$ in the normal state can be usually described by $C = C_e + C_l$, where $C_e = \gamma_n T$ is the electronic contribution, and $C_l(T) = \beta T^3 + \delta T^5$ is the lattice contribution. The dashed curve in figure 2 is the best fitting of the data to this model for $H$ = 0 Oe and $T \leq$ 10 K, yielding the parameters $\gamma_n$ = 10.8(5) mJ/mol K$^2$, $\beta$ = 0.39(3) mJ/mol K$^4$ (corresponding Debye temperature $\Theta_D$ = 246 K), and $\delta$ = 0.2(3) μJ/mol K$^6$. Compared with the matrix 2$H$-TaS$_2$ with $\gamma_n$ = 8.5 mJ/mol K$^2$ and $\beta$ = 0.37 mJ/mol K$^4$ (shown in the table 2) [22], the value of $\gamma_n$ for Cu$_{0.03}$TaS$_2$ is slightly larger, while the value of $\beta$ is almost the same.

With the McMillan formula [23]

$$\lambda = \frac{\mu^* \ln(\frac{1.45T_C}{\Theta_D}) - 1.04}{1.04 + \ln(\frac{1.45T_C}{\Theta_D})(1 - 0.62\mu^*)},$$

the electron-phonon coupling constant $\lambda$ is estimated to be ~ 0.68 by assuming the Coulomb pseudopotential $\mu^*$ = 0.15, which is a typical value of an intermediate coupling BCS superconductor.

The temperature dependence of the $C_e/T$ for $H$ = 0 Oe near the superconducting transition is shown in figure 3. From the obtained ln ($C_e/\gamma_n T_C$) vs $T_C/T$ data shown in the inset of figure 3, the ratio of the gap and the critical temperature is about, $2\Delta/k_B T_C$ = 4.09, significantly larger than the BCS value 3.53 in weak coupling limit [19]. The dashed curve depicted in figure 3 is the theoretical result of the isotropic s-wave BCS gap with $2\Delta/k_B T_C$ = 4.09, in good agreement with the experimental data. The extracted specific heat jump at $T_C$, $\Delta C/\gamma_n T_C$ = 1.64, is also significantly larger than the weak coupling value 1.43, implying intermediate coupling [23]. This value is similar to $\Delta C/\gamma_n T_C$ = 1.68 observed for another dilute TMDC, Cu$_x$TiSe$_2$ [9], although it is markedly less than the observed values 1.9 in 2$H$-TaS$_2$ and 2.1 in 2$H$-NbSe$_2$ [22]. All the determined parameters of the specific heat compared with 2$H$-TaS$_2$, 2$H$-NbSe$_2$, and Cu$_x$TiSe$_2$ are listed in table 2.

*3.3 Calculation and discussions*

According to the experimental X-ray diffraction data, the lattice constants $a$ and $c$ of $Cu_xTaS_2$ are expanded from those of $TaS_2$, $a$ = 0.3310 to 0.3312 nm and $c$=1.2080 to 1.2137 nm, strongly suggesting that Cu is intercalated into the van der Waals gap. In our calculation, the Cu position is assumed to locate at the fractional coordinate (0, 0, 1/2). The DOS near $E_F$ of $Cu_xTaS_2$ at x = 0, 1/16 and 1/8 is shown in figure 4. It is found that the Fermi energy of $2H$-$TaS_2$ is very close to a sharp DOS peak, in agreement with the earlier results [24]. This also resembles to the DOS in CDW $Cu_xTiSe_2$ [25], indicating that these compounds lie in the edge of a spatial charge modulating phase. Thus $2H$-$TaS_2$ is unstable with respect to the transition into the CDW phase. In the homogeneous $2H$-$TaS_2$, the theoretical DOS near the $E_F$, N(0), is about 1.2 states/eV-cell-spin. After Cu intercalation, the theoretical N(0) for $Cu_xTaS_2$ at x = 1/16 is about 1.0 states/eV-cell-spin, which is slightly smaller than that in 2H-$TaS_2$. The slight decrease in N(0) originates from the fact that the Cu intercalation brings more carriers, leading to the increase in $E_F$ and the Fermi energy shifts right from the DOS peak, as shown in figure 4. Meanwhile, the carriers from coppers enlarge the volume of the Fermi surface through entering the CDW gap, as seen in figure 4. We notice that the increase of the Fermi energy was also found in CDW and superconducting $Cu_xTiSe_2$ [26]. Thus in $Cu_xTaS_2$, on the one hand, the Cu intercalation brings more and more carriers into the Fermi surface and diminishes the CDW gap, as we experimentally find that the CDW transition temperature decreases from 78 K in $2H$-$TaS_2$ to 50 K in $Cu_{0.03}TaS_2$ [18]; on the other hand, more carriers can participate the superconducting pairing and raise the $T_C$. Therefore, Cu-intercalated $TaS_2$ suppress the CDW instability, favoring the superconductivity in $Cu_xTaS_2$.

From figure 4, we also found that the single DOS peak in $2H$-$TaS_2$ splits into two peaks in $Cu_xTaS_2$, implying that the rigid band model [10] is not a good approximation for describing the evolution of the electronic structures upon Cu doping. Meanwhile, we plot the TB-LMTO band structures of $Cu_xTaS_2$ for the 2×2×2 supercell. The energy band structure near $E_F$ considerably changes upon intercalation, as shown in figure 5. Compared to the electronic structures of $2H$-$TaS_2$ [24], though Cu 3d bands lie below $E_F$, a fraction of Cu 4s electrons participate in the Fermi surface, leading to obvious changes in the band structure near $E_F$ in $Cu_xTaS_2$. The considerable variations in the DOS, the electronic structures and the existence of an optimal doping in superconducting $Cu_xTaS_2$ [18] demonstrate that the rigid band model is invalid in Cu-intercalated $TaS_2$.

In layered superconducting TMDC, the relationship among the electron-phonon coupling, the superconductivity and the CDW order has been explored for many years and is still under debate. As we see in figure 4 and figure 5, the Cu 4s electrons gained via Cu intercalation fill in the CDW gap and suppress CDW order; this leads $Cu_xTaS_2$ to transit from the CDW phase to the superconducting phase. Recently, Cava et al. [9] reported that Cu-Intercalated $Cu_xTiSe_2$ also undergoes a transition from CDW order to a superconducting phase; the properties of $Cu_xTiSe_2$ in the CDW and the superconducting phases are very similar to the behavior we observe here for Cu-doped $TaS_2$, including the considerable increase in $E_F$ with Cu doping. On the other hand, other authors [27, 28, 29] showed that in $TiS_2$, neither does there exist the CDW order, nor does Cu doping induce a superconducting transition, though the processes of the sample preparation are identical to those for $Cu_xTaS_2$. From these results, one may find a close relationship between the CDW order and the superconductivity in TMDC: strong electron-phonon coupling not only promotes a CDW ground state, but also drives superconducting pairing in layered TMDC. In contrast, the electron-phonon coupling in $TiS_2$ is so weak that neither CDW order nor superconductivity can form [28, 29].

## 4. Conclusion

In summary, the electronic structure and anisotropic superconducting state parameters of $Cu_{0.03}TaS_2$ have been determined. The GL anisotropy, the specific heat jump, the gap ratio and the electron-phonon coupling definitely show that $Cu_{0.03}TaS_2$ is an anisotropic, intermediate-coupling, type-II BCS superconductor. First-principles electronic structure calculations suggest that copper intercalation of $2H$-$TaS_2$ causes a considerable increase of the Fermi surface volume and the carrier density, which suppresses the CDW fluctuations and favors the raise of $T_C$.

**Acknowledgements**

This work was supported by the National Key Basic Research under contract No.2006CB601005, 2007CB925002, and the National Nature Science Foundation of China under contract No.10774146, 10774147, the Bairen Project, and Director's Fund of Hefei Institutes of Physical Science, Chinese Academy of Sciences. Research at the University of Kentucky was supported by the U.S. Dept. Energy Grant No. DE-FG02-97ER45653 and U.S. National Science Foundation Grants Nos. DMR-0240813 and DMR-0552267.

**References**

[1] Wilson J A and Yoffe A D *Adv. Phys*. **18** 193 (1969).

[2] Wilson J A, Di Salvo F J and Mahajan S *Adv. Phys*. **24** 117 (1975).

[3] Hannay N B, Geballe T H, Matthias B T, Andres K, Schmidt P and MacNair D 1965 *Phy. Rev. Lett*. **14** 225

[4] Gamble F R, Di Salvo F J, Klemm R A and Geballe T H 1970 *Science* **168** 568

[5] Dresselhaus M S and Dresselhaus G 2002 *Adv. Phys*. **51** 1

[6] Friend R H and Yoffe A D 1987 *Adv. Phys*. **36** 1

[7] Weller T E, Ellerby M, Saxena S S, Smith R P and Skipper N T 2005 *Nature Phys*. **1** 39

[8] Fang L, Wang Y, Zou P Y, Tang L, Xu Z, Chen H, Dong C, Shan L and Wen H H 2005 *Phys. Rev. B* **72** 014534

[9] Morosan E, Zandbergen H W, Dennis B S, Bos J W G, Onose Y, Klimczuk T, Ramirez A P, Ong N P and Cava R J 2006 *Nature Phys*. **2**, 544 and references therein.

[10] McCanny J V 1979 *J. Phys. C: Solid State Phys*. **12** 3283

[11] Eberhardt W, McGovern I T, Plummer E W and Fisher J E 1980 *Phys. Rev. Lett*. **44**, 200

[12] Harper J M E, Geballe T H and Di Salvo F J 1977 *Phys. Rev. B* **15** 2943

[13] Meyer S F, Howard R E, Stewart G R, Acrivos J V and Geballe T H 1975 *J. Chem. Phys*. **62** 4411

[14] Schlicht A, Lerf A and Biberacher W 1999 *Synthetic Metals* **102** 1483

[15] Lerf A, Sernetz F, Biberacher W and Schöllhorn R 1979 *Mat. Res. Bull*. **14** 797

[16] Fleming R M and Coleman R V 1975 *Phys. Rev. Lett*. **34** 1502

[17] Wagner K E, Morosan E, Hor Y S, Tao J, Zhu Y, Sanders T, McQueen T M, Zandbergen HW,

Williams A J, West D V and Cava R J 2008 *Phys. Rev. B* **78** 104520

[18] Zhu, X D, Sun Y P, Zhu X B, Luo X, Wang B S, Li G, Yang Z R, Song W H and Dai J M 2008 *J. Crystal Growth* **311** 218

[19] Bardeen J, Cooper L N and Schrieffer J R 1957 *Phys. Rev.* **108** 1175

[20] Kunchur M N and Poon S J 1991 *Phys. Rev. B* **43** 2916

[21] Werthamer N R, Helfand E andHohenberg P C 1966 *Phys. Rev.* **147** 295

[22] Meyer S F, Howard R E, Stewart G R, Acrivos J V and Geballe T H 1975 *J. Chem. Phys.* **62** 4411

[23] McMillan W L 1968 *Phys. Rev.* **167** 331

[24] Guo G Y and Liang W Y 1987 *J. Phys. C: Solid State Phys.* **20** 4315

[25] Jeong T and Jarlborg T 2007 *Phys. Rev. B* **76** 153103

[26] Zhao J F, Ou H W, Wu G, Xie B P, Zhang Y, Shen D W, Wei J, Yang L X, Dong J K, Arita M, Namatame H, Taniguchi M, Chen X H and Feng D L 2007 *Phys. Rev. Lett.* **99** 146401

[27] Tazuke Y, Kuwazawa K, Onishi Y and Hashimoto T 1991 *J. Phys. Soc. Jpn.* **60** 2534

[28] Kusawake T, Takahashi Y, Ohshima K and Wey M Y 1999 *J. Phys. Condens. Matter* **11** 6121

[29] Kusawake T, Takahashi Y, Wey M Y and Ohshima K 2001 *J. Phys. Condens. Matter* **13** 9913

**Table 1.** Superconducting parameters for $Cu_{0.03}TaS_2$: critical temperature for superconductivity $T_C$, Sommerfeld coefficient $\gamma_n$, Debye temperature $\Theta_D$, electron-phonon

coupling constant λ, specific heat jump $\Delta C/\gamma_n T_C$, gap ratio $2\Delta/k_B T_C$, upper critical field $H_{C2}(0)$, lower critical field $H_{C1}(0)$ (after demagnetization correction), thermodynamic critical field $H_C(0)$, GL parameters $\kappa(0)$, GL coherence length $\xi_{GL}(0)$, GL penetration depth $\lambda(0)$, GL anisotropy ratio of $H_{C2}$ $\gamma_{anis}$.

| Superconducting State Parameters for $Cu_{0.03}TaS_2$ | | | |
|---|---|---|---|
| | | $H \parallel ab$ | $H \parallel c$ |
| $T_C$ (K) | 4.03 | | |
| $\gamma_n$ (mJ / mol K$^2$) | 10.8(5) | | |
| $\Theta_D$ (K) | 246 | | |
| λ | 0.68 | | |
| $\Delta C/\gamma_n T_C$ | 1.64 | | |
| $2\Delta/k_B T_C$ | 4.09 | | |
| $H_{C2}(0)$ (T) | | 9.16 | 1.8 |
| $H_{C1}(0)$ (Oe) | | 40 | 135 |
| $H_C(0)$ (T) | ~ 0.1 | | |
| $\kappa(0)$ | | 69.7 | 13.1 |
| $\xi_{GL}(0)$ (nm) | | 13.5 | 2.65 |
| $\lambda(0)$ (nm) | | 177 | 983 |
| $\gamma_{anis}$ | 5.1 | | |

Tab. 1 X.D. Zhu et al.

**Table 2.** The superconducting transition temperature ($T_C$) and specific heat parameters for $Cu_{0.03}TaS_2$, compared with those of 2$H$-TaS$_2$, Py$_{1/2}$TaS$_2$, 2$H$-NbSe$_2$, and Cu$_x$TiSe$_2$.

| Compound | $T_C$ (K) | $\gamma_n$ (mJ / mol K$^2$) | $\beta$ (mJ / mol K$^4$) | $\Delta C/\gamma_n T_C$ | Ref. |
| --- | --- | --- | --- | --- | --- |
| Cu$_{0.03}$TaS$_2$ | 4.03 | 10.8(5) | 0.39(3) | 1.64 | This work |
| 2$H$-TaS$_2$ | 0.8 | 8.5 | 0.37 | 1.9 | [17] |
| Py$_{1/2}$TaS$_2$ | 3.5 | 9.1 | 2.32 | 0.96 | [17] |
| 2$H$-NbSe$_2$ | 7.1 | 16.5 | 0.53 | 1.73 | [17] |
| Cu$_x$TiSe$_2$ | 4.1 | 4.3 | … | 1.68 | [8] |

Tab. 2 X.D.Zhu et al.

**Figures:**

Figure 1

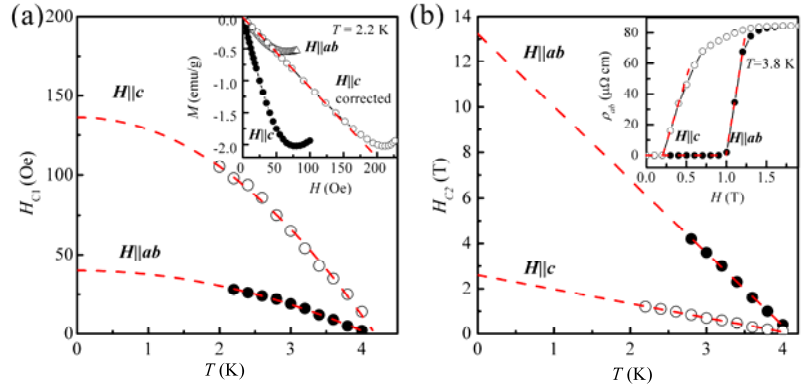

**Figure 1.** (a) Temperature ($T$) dependence of the lower critical field ($H_{C1}$) for $Cu_{0.03}TaS_2$. The dashed line shows the fitted curve. The inset shows the magnetization as a function of applied field (*M-H* curves) measured at $T$ = 2.2 K, for ***H** || **ab***, ***H** || **c*** and ***H** || **c*** with demagnetization effect correction. (b) Temperature dependence of the upper critical field $H_{C2}$ for $Cu_{0.03}TaS_2$. The dashed line shows the fitted linear curve. The inset shows in-plane resistivity ($\rho_{ab}$) as a function of applied field both for ***H** || **ab*** and ***H** || **c*** measured at $T$ = 3.8 K.

Figure 2

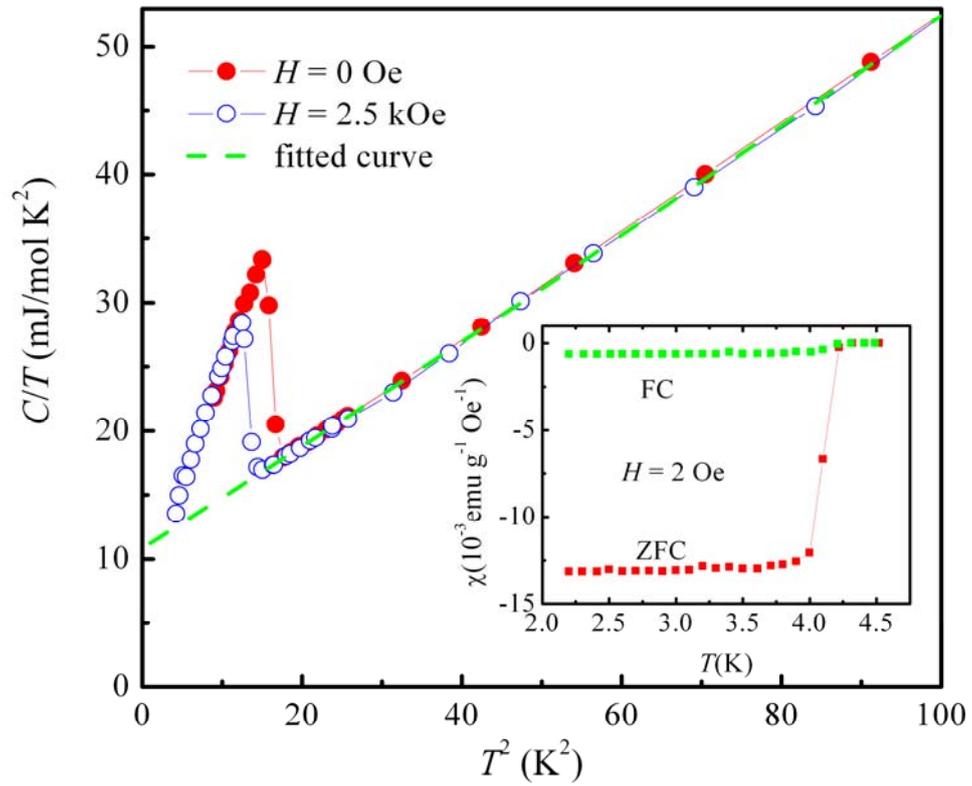

**Figure 2.** Specific heat divided by temperature ($C/T$) as a function of $T^2$ for $Cu_{0.03}TaS_2$ measured at $H = 0$ (solid circles, ●) and $H = 2.5$ kOe (open circles, ○). The dashed line represents the best-fit curve. The inset shows the temperature dependence of the dc magnetic susceptibilities with $H = 2$ Oe applied parallel to *ab*-plane under zero-field-cooled (ZFC) and field-cooled (FC) conditions.

Figure 3

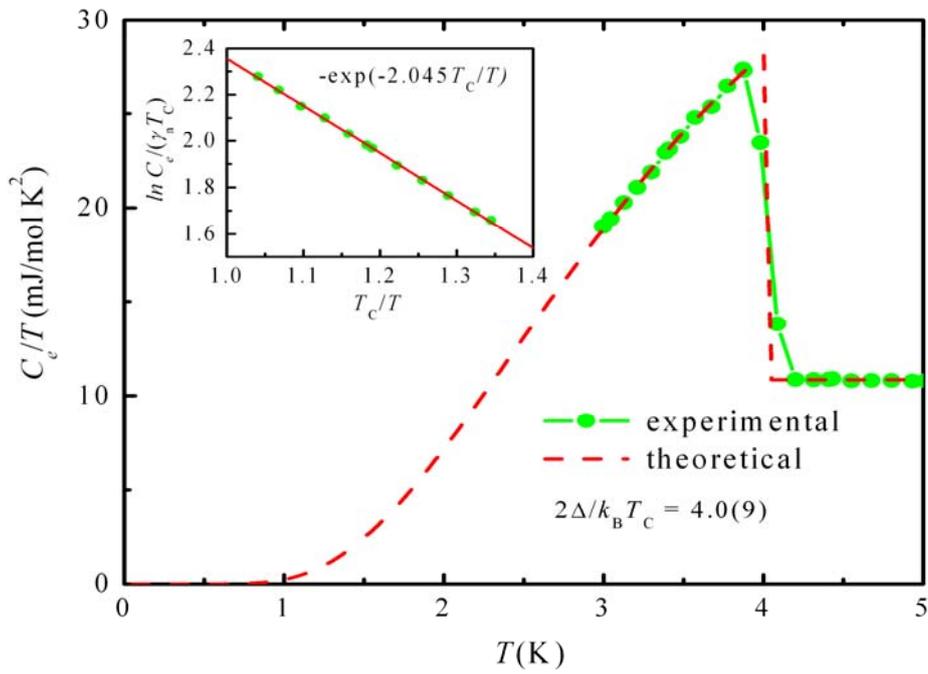

**Figure 3.** Temperature dependence of the electronic specific heat divided by temperature, $C_e/T$ at $H$ = 0 Oe for $Cu_{0.03}TaS_2$. The solid line shows $C_e/T$ calculated by assuming an isotropic s-wave BCS gap with $2\Delta/k_BT_C$ = 4.09.

Figure 4

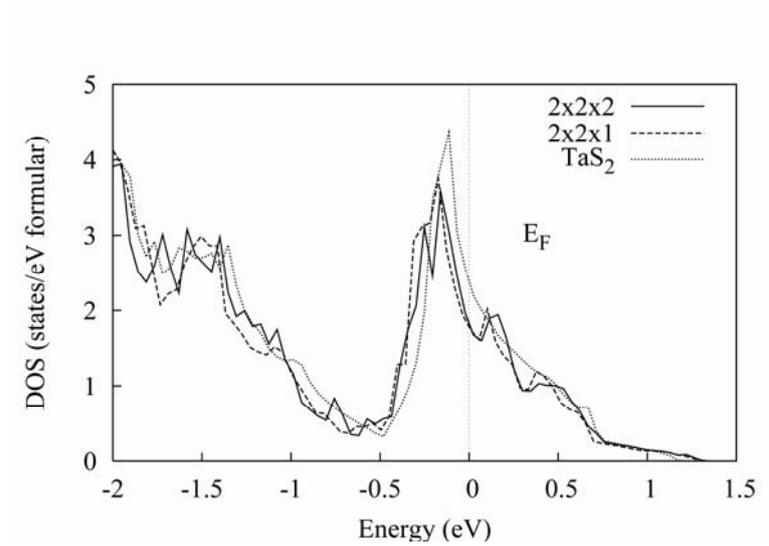

**Figure 4.** The density of states (DOS) in $Cu_xTaS_2$ in normal state for the supercell 2×2×2 (solid line) and the supercell 2×2×1 (dash line). The DOS of undoped $2H$-$TaS_2$ is also plotted for comparison (dot line).

Figure 5

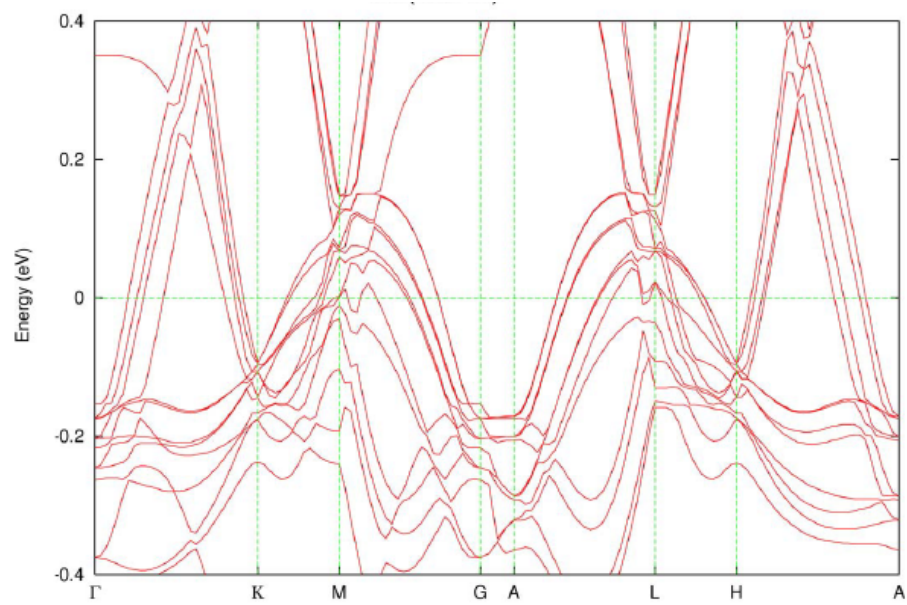

**Figure 5.** The energy band structures of $Cu_xTaS_2$ by the TB-LMTO-ASA method for the supercell 2×2×2, corresponding to Cu concentration x = 1/16. Details of the electronic structure of undoped $2H$-$TaS_2$ can be found in Ref. [24] for comparison.